# Quantitative measurement of the magnetic moment of an individual magnetic nanoparticle by magnetic force microscopy.

K.-F. Braun[1], S. Sievers[1], D. Eberbeck[2], S. Gustafsson[3], E. Olsson[3], H. W. Schumacher[1], U. Siegner[1]

[1]*Physikalisch-Technische Bundesanstalt, Bundesallee 100, 38116 Braunschweig, Germany*

[2]*Physikalisch-Technische Bundesanstalt, Abbestraße 2-12, 10587 Berlin, Germany*

[3]*Department of Applied Physics, Chalmers University of Technology, 41296 Gothenburg, Sweden*

We demonstrate the quantitative measurement of the magnetization of *individual* magnetic nanoparticles (MNP) using a magnetic force microscope (MFM). The quantitative measurement is realized by calibration of the MFM signal using an MNP reference sample with traceably determined magnetization. A resolution of the magnetic moment of the order of $10^{-18}$ Am$^2$ under ambient conditions is demonstrated which is presently limited by the tip's magnetic moment and the noise level of the instrument. The calibration scheme can be applied to practically any MFM and tip thus allowing a wide range of future applications e.g. in nanomagnetism and biotechnology.

Magnetic nanoparticles (MNP) show potential use for a wide range of applications for example in biomedicine [1] and for data storage [2]. For research purposes as well as for quality control, a precise characterization of the magnetic properties of the MNPs is essential. However, standard characterization techniques like SQUID magnetometry only allow for the measurement of integral properties of ensembles of MNPs. A direct characterization of individual particles is only possible by microscopy techniques.

Due to its high spatial resolution, magnetic force microscopy (MFM) is a powerful tool for imaging magnetic nanostructures. MFM is a stray field sensitive technique with a resolution down to 10 nm. However, a quantitative interpretation of the measured stray field data is not straight forward. The standard approach for the quantitative characterization of small structures is the point probe approximation [3-5]. However, since the point probe approach disregards the non-local character of the MFM tip magnetization, the approximation is inadequate for patterns with dimensions comparable to the tip dimensions.

In this paper it is shown that a calibration of MFM tips can be obtained for the quantitative measurement of the magnetic moment of spherical nanoparticles. No assumption regarding the tip geometry is required since the stray field of a homogeneously magnetized sphere equals the stray field of a point dipole positioned in the centre of the MNP. This calibration scheme is based on an MNP reference sample, which provides traceability to the SI units for the measurement of magnetic moments of *individual* MNPs as small as $10^{-18}$ Am$^2$.

In magnetic force microscopy, the tip scans over the sample at a given lift height $h$ and the frequency shift $\Delta f$ of the oscillating MFM cantilever is recorded. The frequency shift $\Delta f$ can be calculated from the force $\boldsymbol{F}$ that is acting on the magnetic tip in the stray field $\boldsymbol{H}$ of the sample as $\Delta f = Q/k \cdot d/dz \, F_z = Q/k \cdot d^2/dz^2 \, E_{\text{tip-sample}}$. Here, $k$ and $Q$ are the spring constant and the quality factor of the oscillating cantilever, respectively, $F_Z$ is the component of $\boldsymbol{F}$ perpendicular to the sample surface, and $E_{\text{tip-sample}}$ is the interaction energy between the magnetic stray field of the MNP and the tip. In general, the magnetic coating of an MFM tip has a finite spatial extent. Hence, $E_{\text{tip-sample}}$ can be expressed in terms of a convolution of the tip magnetization $\boldsymbol{M}_{\text{tip}}$ and the sample stray field $\boldsymbol{H}$, which reads for a tip whose apex is at the position $\boldsymbol{r} = (x, y, z)$:

$$E_{tip-sample}(\mathbf{r}) = \int_{tip} \mathbf{M}_{tip}(\mathbf{r}' - \mathbf{r}) \cdot \mathbf{H}(\mathbf{r}') d\mathbf{r}' \quad (1)$$

Now, we focus on single domain MNPs, which can be modeled as magnetic nanospheres with saturation magnetization $M_s$ and volume $V = 1/6 \, \pi d^3$, with $d$ being the diameter of the MNP. For this geometry the stray field $\boldsymbol{H}$ is equal to the stray field of a magnetic dipole which is positioned in the center of the sphere [6]. The absolute value $m$ of the dipole moment $\boldsymbol{m}$ is then given by $m = M_s \cdot V = M_s/6\pi d^3$ and the stray field of an MNP that is located at $\boldsymbol{r}' = 0$ is given by:

$$\mathbf{H}(\mathbf{r}') = \frac{(\mathbf{m} \cdot \mathbf{r}')\mathbf{r}' - r'^2\mathbf{m}}{r'^5} \quad (2)$$

If the magnetic anisotropy of a nanoparticle is sufficiently small, the stray field emerging from the magnetic tip is sufficient to fully align the nanoparticle magnetization[7] as sketched in Fig. 1 a). Since for the most common MFM tips the stray field underneath the tip is oriented perpendicular to the x-y scanning plane, also the magnetization of the nanoparticle is aligned perpendicular when the MFM tip is located at a lift height $h$ above the centre position of the nanoparticle, i.e. at $\boldsymbol{r} = (0,0,z) = (0,0,h)$. At this specific position, the particle magnetization $\boldsymbol{m}$ is given by $\boldsymbol{m}=m\cdot\boldsymbol{z}$, with $\boldsymbol{z}$ the unit vector in z-direction. Hence, the frequency shift $\Delta f$ over the centre of the MNP can be calculated as

$$\Delta f\left(\mathbf{r}=(0,0,h)\right)=m\int_{tip}\frac{Q}{k}\cdot\frac{d^2}{dz^2}\mathbf{M}_{tip}\left(\mathbf{r}'-h\cdot\mathbf{z}\right)\cdot\frac{(\mathbf{z}\cdot\mathbf{r}')\mathbf{r}'-r'^2\mathbf{z}}{r'^5}d\mathbf{r}'.\ (3)$$

The integral term becomes a constant that only depends on the magnetic properties of the magnetic probe. For a given tip height $h$ it therefore represents a tip dependent proportionality constant $1/c(h)$ connecting the magnetic moment $m$ of the spherical nanoparticle and the measured MFM frequency shift by $\Delta f=m/c(h)=\pi M_s d^3/6c(h)$. As a consequence, for spherical MNPs a calibration of the magnetic tip can be achieved by measuring the MFM signal of a set of nanoparticles with known magnetic moment.

In the following, the realization and characterization of such an MNP based reference sample for MFM calibration is described. A suitable MNP reference sample has to fulfill the following requirements: (i) the MNPs do not agglomerate and (ii) the magnetization of the MNPs is well known. We selected commercial magnetite nanoparticles with 20 nm nominal diameter, in the following referred to as SHP 20 ([a]). A sample of well separated MNPs was prepared by pouring the particles in solution onto a silicon substrate which is exposed to a vertical magnetic field ($\approx$ 500 mT). Thereby the particles are magnetically aligned and repel each other which prevents particle agglomeration during the drying process. The MNP's size distribution was first determined by transmission electron microscopy (TEM) (Fig. 1b). The resulting mean particle diameter is $d_{\text{TEM}}=(18.7\pm3)$ nm. To traceably determine the saturation magnetization $M_S$ of the reference MNPs we first measured the total magnetic moment of a small sample volume of the MNP suspension by SQUID magnetometry. Then, the iron content of the same sample volume was determined by titration using prussian blue staining. From the total magnetic moment and the magnetite volume derived by titration the saturation magnetization was determined to be $M_S=(250\pm10)$ kA/m at 293 K. The measured value of $M_S$ allows for a calculation of the magnetic moment of the SHP20 MNPs for a given particle diameter using the relation $m=M_S \cdot V=M_s/6\pi d^3$. Note that in these considerations possible effects resulting from a non-magnetic shell are not accounted for.

A SHP20 reference sample, prepared as described above, was then employed to calibrate the signal of commercial MFM cantilevers ([b]). Atomic force microscopy (AFM) and MFM was performed using an SIS instrument working in a self excitation mode. The tip scans the sample at a constant lift height $h$ with respect to the sample surface. For the calibration in a first step an AFM topography image was recorded as shown in Fig. 2a. The height and thus the diameter of the particles was determined by fitting a two dimensional Gaussian function to each scanned particle (see Fig. 2c, red solid line). The theoretical AFM topography curve of a spherical nanoparticle (Fig. 2c, black dashed line) can be described by a convolution of two semicircles describing the spherical particle of diameter $d$ and the tip with a certain tip curvature radius. In contrast a Gaussian fit (red solid line) overestimates the amplitude by a factor of about 1.08. This factor is practically constant over the range of particles used here. Therefore, for ease of computation, we used the two dimensional Gaussian fit and corrected the resulting amplitudes by this factor to determine the particle diameter $d$. For the AFM image shown in 2(a) the resulting mean particle diameter is $d_{AFM}=(17.1\pm2.7)$ nm, in good agreement with the TEM analysis.

In a second step, the corresponding MFM images were taken at a constant lift height of $h=60$ nm (Fig. 2b). In the MFM image the MNP appear as a depression consistent with the concept of a particle that is magnetized by the magnetic stray field of the tip [7]. As described above the calibration has to be carried out when the tip is positioned directly over the MNP. Only here, the moment of the MNP is aligned perpendicular by the tip field and the frequency shift $\Delta f$ of the MFM signal is governed by the relation described in equation 3. Note further that in this position the maximum frequency shift $\Delta f$ for each particle is detected. We determine the maximum frequency shift at the centre position again by a two dimensional Gaussian function. A typical example of such fit for one MNP is shown in Fig. 2b (red solid line). The fitted Gaussian function well describes the measured MFM signal. The exact functional form of the frequency signal of a magnetic cantilever crossing a magnetic nanoparticle could in principle be calculated under the assumption of the magnetic moment being aligned parallel to the momentary stray field of the tip. However, neither the magnetization of the tip nor the stray field are known. Fitting the measured MFM signal using the simple point dipole tip model results in a too narrow linewidth compared to the measured data (Fig. 2d, black dashed line). Hence the tip stray field can not be suitably described by a point dipole which gives further evidence for the need of a calibration scheme beyond a simple point dipole approximation.

The heights $d$ and MFM signals $\Delta f$ of 73 particles from Fig. 2a and 2b have been determined. In Fig. 2e the measured frequency shift is plotted as a function of $d^3$. The displayed data basically shows a linear increase, however considerably scattered. Fitting the data by a linear regression allows to derive the tip calibration factor $c(h)$ using $c(h)^{-1}=6\Delta f/(\pi M_s d^3)$.

The resulting tip calibration factor is $c^{-1}(h=60nm)=(1.38\pm0.36)\dfrac{Hz}{A\cdot nm^2}$, and hence

$c(h=60nm)=(0.72\pm0.19)\dfrac{A\cdot nm^2}{Hz}$. The interception of the linear regression with the y axis is zero within statistical error estimates $b$=(-0.17±0.22) as expected, hence no significant contribution of a non magnetic shell of the MNP is found. The derived calibration factor $c(h)$ thus relates the MFM signal for the given MFM tip and lift height to the absolute value $m$ of a magnetic moment of a specific MNP. Hence the calibrated MFM tip operating at the given tip lift height $h$ can be used to *traceably* measure the magnetic moment of any other unknown MNP that fulfills the following two conditions: (a) the MNP has an approximately spherical shape, and (b) the anisotropy is sufficiently weak so that the magnetic moment $m$ is aligned by the stray field of the tip. Note that the calibration of another tip of the same type resulted in a calibration factor of $c(h = 60nm) = (2.37 \pm 0.45) \frac{A \cdot nm^2}{Hz}$, reflecting the differences in the mechanical und magnetic properties of nominally equal tips.

As mentioned before the values plotted in Fig. 2c significantly scatter around the linear regression. One reason for such scatter could be a drift of the nominally constant lift height $h$ due to piezo creep and piezo hysteresis during the MFM scan. As a consequence particles of the same size would show a different frequency shift at different positions of the scan resulting in scattered data. Additionally, thermal activation can induce a fluctuation of the magnetization around the axis defined by the tip field and, thus, gives rise to scattered data. Note, however, that such fluctuation of the alignment out of the field axis will result in an underestimation of the frequency shift for the given sample size. A similar effect could occur if a non-negligible magnetic anisotropy is present in some of the particles. Also such anisotropy would inhibit a full alignment of the magnetic moment in the stray field of the tip and thus would result in an underestimated frequency shift. The MFM tip scans at a constant lift height $h$ with respect to the sample surface. The diameters of the measured particles show values ranging from 9.2 nm to 22.7 nm. Consequently during the scan with the nominal lift height of 60 nm the distance between the tip apex and the center point of the particles is not constant, but varies from 55.4 nm to 48.7 nm, i.e. by $\Delta h$=6.7 nm. This causes a systematic error $\Delta c$ of the calibration factor of about 10 %, as can be estimated from the height dependence of the calibration factor $c$. In Fig. 2f the calibration factor $c(h)$ derived for the same tip and three different lift heights of $h = 50, 60,$ and 70 nm is plotted as a function of $h$.

The sensitivity of the tip decays with increasing distance as expected. The line in Fig. 2f serves as a guide to the eye.

The calibrated tip characterized by the data of Fig. 2 was used to characterize a different magnetite MNP sample with a nominal particle diameter of 30 nm ([c]), whose magnetic properties were not known a priori. The MFM measurement is shown in the inset of Fig. 3. A quantitative analysis of the magnetic moment of one of the MNPs in Fig. 3 a) is exemplarily shown in the Fig. 3 b). It shows a scan of the MFM frequency shift $\Delta f$ measured along the line in Fig. 3 a) as a function of the tip position during the scan. The MFM image was measured at a lift height of 60 nm. The maximum frequency shift is again derived from a two dimensional Gaussian fit of the resulting bell curve (red solid line). The offset of the Gaussian was fixed since the zero line was independently determined from the background signal of the MFM image. From the peak height of the frequency shift $\Delta f$=(1.155±0.065) Hz and using the tip calibration factor $c(h$=60nm) given above the absolute value of the magnetic moment of the MNP under consideration is determined to be $m$=(0.84±0.27) Anm$^2$. The measurement uncertainty of this traceable measurement results from the uncertainty of the frequency measurement (i.e. the system noise) and the uncertainty of the calibration factor $c$. Note that, in principle, similar analysis of the frequency shift of the other MNPs would yield the distribution of the quantitatively measured magnetic moments of the set of MNPs under consideration.

For our present measurement setup using the SIS instrument working in a self excitation mode the value of magnetic moment to be reliably resolved is limited by the resolution of our instrument corresponding to a frequency shift of 0.5 Hz. With a tip with a calibration factor of $c(h = 60nm) = (0.72 \pm 0.19) \frac{A \cdot nm^2}{Hz}$, as it has been determined above, the minimum magnetic moment that can be resolved is 0.36 Anm$^2$. The limitation is mainly due to the noise in the frequency measurement resulting from mechanical and electrical sources. Hence using an improved setup using e.g. a low noise instrument a higher magnetic moment resolution is possible.

To conclude, we presented a technique for the traceable calibration of MFM tips that allows for a quantitative measurement of the magnetic moments of individual magnetic nanoparticles. The resolution of the technique is only limited by the intrinsic noise of the MFM under regard. The calibration is based on the characterization of a reference sample consisting of well characterized magnetic nanoparticles. The magnetic parameters of the reference sample were determined by SQUID measurements and thereby assuring traceability of the MFM calibration to the SI system of units.

This work is supported through the Federal Ministry of Education and Research under the grant number 13N9149 and 13N9150.

**Footnotes:**

([a]) Type SHP-20-0010 from Ocean Nano Tech LLC

([b]) Nanosensors, PPP-MFMR

([c]) SHP-30-0010 from Ocean Nano Tech LLC

**Figure Captions:**

*Fig.1: (a) The particles magnetic moment is aligned with the magnetic field below the tip. (b)The inset shows a TEM picture of magnetic nanoparticles of the sample SHP-20. The plot shows the core diameter distribution estimated from the TEM image.*

*Fig.2: (a) AFM image and (b) MFM image recorded at a liftheight of 60 nm of the sample SHP-20 (same sample area, 2 x 2 µm). (c) and (d) show linescans across a nanoparticle (see text). (e)Plot of the calculated values of MFM signal versus the cubed diameter. The solid line shows the linear fit. (d) Calibration factor as a function of the liftheight. The solid line is a guide to the eye.*

*Fig.3: MFM image (a) and linescan through a single MNP(b). The linescan is evaluated in SI units of the magnetic moment. The data have been smoothed and the solid line is a Gaussian profile.*

**Figure 1:**

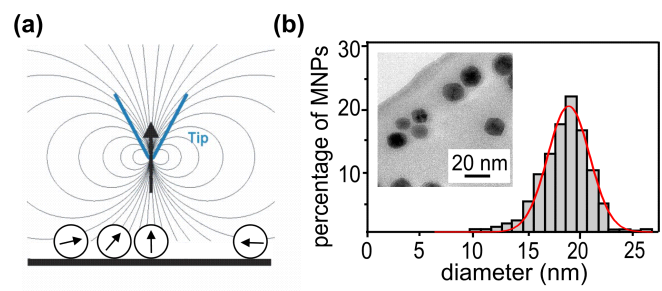

**Figure 2:**

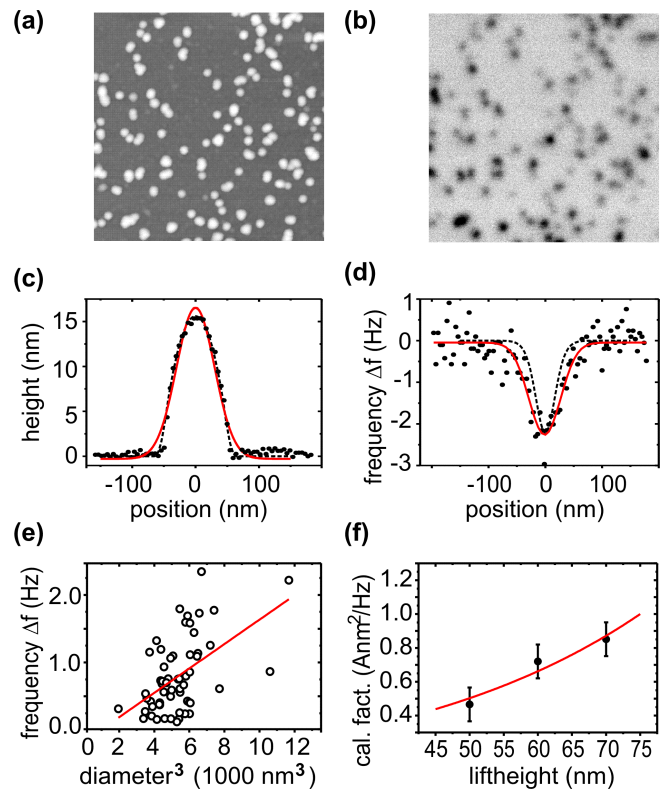

**Figure 3:**

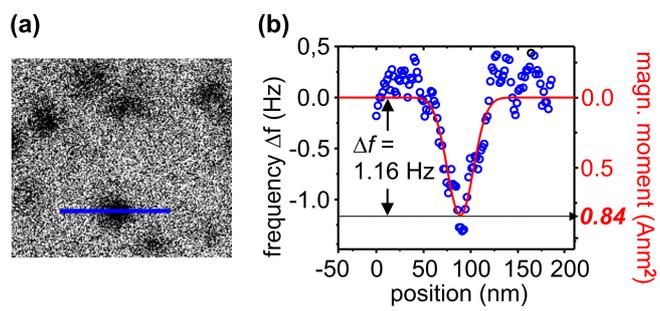